\documentclass[useAMS,usenatbib]{mn2e}
\usepackage{graphicx}
\usepackage{natbib}
\def\Xback{\mbox{erg s$^{-1}$ cm$^{-2}$ deg$^{-2}$}}
\def\Xflux{\mbox{erg s$^{-1}$ cm$^{-2}$}}
\newcommand{\msun}{\,{\rm M_\odot}}

\title[Constraints on the Accretion History of Massive Black Holes from Faint X-ray Counts]{Constraints on the Accretion History of Massive Black Holes from Faint X-ray Counts}
\author[M. Volonteri, R. Salvaterra \& F. Haardt]{Marta Volonteri$^{1}$, Ruben Salvaterra$^{2}$ \& Francesco Haardt$^{2}$\\
\footnotemark[1] 
$^{1}$Institute of Astronomy, Madingley Road, Cambridge CB3 0HA, UK\\
$^{2}$Dipartimento di Fisica e Matematica, Universit\'a dell'Insubria, Via Valleggio 11, 22100 Como, Italy}

\begin{document}

\maketitle

\begin{abstract}
We investigate how hierarchical models for the co-evolution of the massive black hole (MBH) and AGN population
can reproduce the observed faint X-ray counts.  We find that the main variable influencing the theoretical predictions 
is the Eddington ratio of accreting sources.  We compare three different models proposed for the evolution
of AGN Eddington ratio, $f_{\rm Edd}$: constant $f_{\rm Edd}=1$,   $f_{\rm Edd}$ decreasing with redshift, and
$f_{\rm Edd}$ depending on the AGN luminosity, as suggested by simulations of galactic mergers including MBHs and AGN feedback.  We follow the full assembly of MBHs and host halos from early times to the present in a 
$\Lambda$CDM cosmology. AGN activity is triggered by halo major mergers and MBHs accrete mass until they satisfy 
the observed correlation with velocity dispersion.  We find that all three models can reproduce fairly well the total 
faint X-ray counts. The redshift distribution is however poorly matched in the first two models. 
The Eddington ratios suggested by merger simulations predicts no turn-off of the faint end of the AGN optical luminosity function 
at redshifts $z\ga1$, down to very low luminosity. 
\end{abstract}

\section{Introduction}
Several hierarchical models for the evolution of the MBH and AGN populations \citep[see,e.g.,][]{haehnelt1993, haiman2000, hatziminaoglou2001,  wyithe2003, Cattaneo1999, kauffmann2000, cavaliere2000, VMH, Granatoetal, lapi2006} 
 have proved successful in reproducing 
the AGN optical luminosity function (OLF) in a large redshift range ($1\la z\la6$). 
Typically, these models assume that AGN activity is triggered by major mergers.  Galactic interactions  trigger gas inflows, and the cold gas may be eventually driven into the very inner regions, fueling an accretion episode and the growth of the nuclear MBH. Hydrodynamic simulations of major mergers have 
shown that a significant fraction of the gas in interacting galaxies falls to the center of the merged system \citep{mihos94, mihos96}: the cold gas may be eventually driven into the very inner regions, fueling an accretion episode
and the growth of the nuclear BH. This last year has been especially exciting, as the first high resolution simulations of  galactic mergers including BHs and AGN feedback, showed that the merger scenario  is generally correct \citep{Springel2005,dimatteo2005}. In hierarchical models of galaxy formation major mergers are responsible for forming bulges  and elliptical galaxies. Support for 
merger driven activity therefore comes from the observed  correlation between bulge luminosity - 
or stellar velocity dispersion - and black hole mass, suggesting a  single mechanism for 
assembling black holes and forming spheroids in galaxy halos \citep{magorrian1998, Marconietal2004, Gebhardt2000, ferrarese2000}.

Notwithstanding the success at high redshift, hierarchical models struggle to match the AGN OLF at low redshift ($z\la1$) by overpredicting the bright end,  and underpredicting the faint end of the OLF, as the decrease of the halo merger rate with time 
is less dramatic than the observed fall of the AGN population. The overprediction of bright AGN can be imputed to inefficient cooling 
in large halos. Imposing an upper limit to the AGN host halo mass ($\sim10^{13.5} M_{\odot}$, \cite{wyithe2003, Marulli2006}) in fact significantly improves the match at the bright end of the OLF. 

The under-abundance of faint AGN can be instead attributed to the assumption, common to most models, of very efficient accretion, 
at rates close to the Eddington rate.  \cite{merloni2003} and \cite{merloni2004} have shown that low-redshift AGN are probably accreting inefficiently, i.e. both at an accretion rate much smaller than the Eddington rate and with a low
radiative efficiency. These considerations suggest that successful predictions for the evolution of AGN luminosity should
include more sophisticated models for accretion. A first step in this direction can be taken by considering the results of 
the recent merger simulations, which track also accretion on a central MBH. Although these simulations lack the necessary 
resolution for resolving the accretion process in the vicinity of the MBH, empirical models \citep{hopkins2005a}, based on coupling results from the above simulations with the observed LF in the hard X-ray band (HXLF), have been shown to reproduce simulaneously several optical and X-ray observations.  The \cite{hopkins2005a} empirical models, however, are not embedded in a cosmological evolutionary framework, that is they derive the MBH population properties at a given time, but not how the population of black holes at an earlier time evolves into the MBHs present at a later time. 
From the observed HXLF \cite{hopkins2005a} derive the rate at which AGN of a given luminosity at the peak of activity must be created. This information is then used as a proxy for the galaxy merger rate which should provide the boundary conditions for determining the evolution of the MBH population.  

We here couple the predictions from \cite{hopkins2005a} with the merger rate expected in the currently favoured cold dark matter scenario. The main novelties of the present investigation with respect to \cite{hopkins2005a}, are therefore that (i) the rate of mergers, which trigger AGN activity, is directly derived in the currently favoured cold dark matter (CDM) cosmology. In principle, the empirical merger rate derived by \cite{hopkins2005a} is not granted to correspond  to the CDM one.  Also, (ii), we grow MBHs in a self-consistent way, that is we trace the whole accretion history of MBHs from early times to the present, requiring continuity in the population. 

Our aim is to investigate here to which extent hierarchical models, coupled with the prescriptions based on the above simulations (\S1), can reproduce the low-redshift evolution of AGN. We show that reproducing the HXLF is a necessary, but not sufficient, condition for matching the redshift distribution of faint X-ray counts (\S2). We identify the redshift distribution of faint X-ray counts as the most sensitive observational result to discriminate between models (\S3, \S4). Finally, in \S5 we summarize the results and discuss their implications.

\section{Formation of massive black holes and growth by mass accretion}
In our framework pregalactic `seed' holes form at early times. In most of our calculations we follow \cite{VHM,Volonterietal2005}, assuming that seed MBHs form with intermediate masses ($m_{\rm seed}\lid 600\,\msun$) in halos collapsing at $z=20$ from rare 3.5-$\sigma$ peaks of the primordial density field \citep{MadauRees2001}  as end-product of the very first generation of stars. 
The assumed `bias' assures that almost all halos above $10^{11}\,\msun$ actually host a BH at all epochs. 
We also check the influence of the initial conditions by considering seed MBH formation as in \cite{kous}. In this model seed MBH  form from the low angular momentum tail of material in halos with efficient gas cooling.   In first approximation, seed MBH form in halos with mass above the threshold $M_H\simeq10^7\msun (1+z/18)^{-3/2}$, with a mass $m_{\rm seed}\simeq5\times10^4\msun(M_H/10^7\msun)(1+z/18)^{3/2}$ \citep{kous}. We have dropped here the dependency on the halo spin parameter and gas fraction, as we are not interested in the detailed seed formation process, but only  in testing an alternative model for seed formation which predicts much larger seed masses. 

Nuclear activity is triggered by halo mergers: in each major merger the hole in the more massive halo accretes gas until its mass scales 
with the fifth power of the circular velocity of the host halo with a normalization which reproduces the observed local correlation between MBH mass and velocity dispersion ($m_{\rm BH}-\sigma_*$ relation).  

The rate at which mass is accreted scales with the Eddington rate for the MBH. In model I the accretion rate is set exactly to be the 
Eddington rate. Defining the Eddington ratio as $f_{\rm Edd}=\dot {M}/\dot{M}_{\rm Edd}$, model I has a constant $f_{\rm Edd}=1$. 
\cite{Shankar2004} suggest that if the Eddington ratio evolves with redshift  the MBH mass function derived from a deconvolution of the AGN LF agrees better with the local MBH  mass function \citep{aller2002, Shankar2004, Marconietal2004}. 
\cite{Shankar2004} suggest the following parameterization:
\begin{equation}
\label{eq:shankar1}
f_{\rm Edd}(z)=\left\{
\begin{array}{ll}
f_{\rm Edd,0}       &  z\geqslant3   \\
f_{\rm Edd,0}[(1+z)/4]^{1.4}      &   z<3
\end{array}
\right.
\end{equation}
with $f_{\rm Edd,0}=0.3$. Face value,  Equation \ref{eq:shankar1} underpredicts the LFs of AGN at high redshift, in our framework. As we start from small high redshift seeds, for our model II we modify Equation \ref{eq:shankar1} as follows:

\begin{equation}
\label{eq:shankar2 }
f_{\rm Edd}(z)=\left\{
\begin{array}{ll}
1  &  z\geqslant6 \\
0.078(1+z)^2-0.623(1+z)+1.545       & 3\lid z<6   \\
f_{\rm Edd,0}[(1+z)/4]^{1.4}      &   z<3,
\end{array}
\right. 
\end{equation}
where the quadratic form smoothly joins the $z<3$ and $z>6$ functional forms.

Finally, we test models where the Eddington ratio scales with the AGN luminosity. In model IIIa we parameterize the accretion rate following \cite{hopkins2005b}. The results of simulations are presented in \cite{hopkins2005b} in terms of AGN luminosity. As our main variable is the black hole mass, we introduce some simplifications to the model, but the general trend is preserved (see \cite{hopkins2006} for a thorough discussion). 
The time spent by a given AGN per logarithmic interval is approximated by \cite{hopkins2005b} as:
\begin{equation}
\label{eq:dtdlogL }
\frac{{d}t}{{d}L}=|\alpha|t_Q\, L^{-1}\, \left(\frac{L}{10^9L_\odot}\right)^\alpha,
\end{equation}
where $t_Q\simeq10^9$ yr, and $\alpha=-0.95+0.32\log(L_{\rm peak}/10^{12} L_\odot)$. Here $L_{\rm peak}$ is the luminosity of the AGN at the peak of its activity.  \cite{hopkins2006} show that approximating $L_{\rm peak}$ with the Eddington luminosity of the MBH at its final mass (i.e., when it sets on the $m_{\rm BH}-\sigma_*$ relation) the difference in their results is very small. 
If we write the accretion rate in terms of the time-varying Eddington rate:
$\dot {M}=f_{\rm Edd}(t)\dot{M}_{\rm Edd}=f_{\rm Edd}(t) m_{\rm BH}/t_{\rm Edd}$ ($t_{\rm Edd}=0.45\,{\rm Gyr} $),
the AGN luminosity can be written as $L=\epsilon f_{\rm Edd}(t)  \dot{M}_{\rm Edd} c^2$, where $\epsilon$ is the radiative efficiency\footnote{We determine the radiative efficiency self-consistently tracking the evolution of black hole spins throughout our calculations \citep{Volonterietal2005}. We adopt  an upper limit to the radiative efficiency of $\epsilon=0.16$, as this corresponds, adopting the standard conversion 
for accretion from a thin disc, to a maximum spin parameter of the BH $\hat a=0.9$. This value was chosen in agreement with \cite{Gammieetal2004} simulations, which suggest that the maximum spin MBHs can achieve by coupling with discs in magneto-hydrodynamical simulations is $\hat a\simeq 0.9$.}.  
Differentiating with respect to the Eddington ratio, we can write a simple differential equation for $\dot f_{\rm Edd}(t)$:
\begin{equation}
\label{eq:doteddratio }
\frac{{d}f_{\rm Edd}(t)}{{d}t}=\frac{ f_{\rm Edd}^{1-\alpha}(t) }{|\alpha| t_Q}\left(\frac{\epsilon \dot{M}_{\rm Edd} c^2}{10^9L_\odot}\right)^{-\alpha}.
\end{equation}
Solving this equation gives us the instantaneous Eddington ratio for a given MBH at a given time, and 
we self-consistently grow the MBH mass:
\begin{equation}
\label{eq:bhgrowth }
M(t+\Delta t)=M(t)\,\exp\left(\int_{\Delta t}\frac{dt}{t_{\rm Edd}} f_{\rm Edd}(t)\frac{1-\epsilon}{\epsilon}\right).
\end{equation}
We define a lower limit to the Eddington ratio $f_{\rm Edd}=10^{-3}$. 

As will be discussed in Section  \S4, we also consider a modification of model IIIa, where we include a much stronger dependence of $f_{\rm Edd}$ on
the galaxy velocity dispersion, in practice we modify the exponent $\alpha$ in Equation \ref{eq:dtdlogL } as follows:
\begin{equation}
\label{eq:doteddalpha}
\alpha=-0.5\left(\frac{V_c}{320}\right)^2+1.5\left(\frac{V_c}{320}\right)^{1/3}\log_{10}\left(1.46\frac{V_c}{320}\right).
\end{equation}
Although not physically motivated, as Equation \ref{eq:dtdlogL }
 is not either,  Equation \ref{eq:doteddalpha} was inspired by the trend in 
accretion rates shown by \cite{dimatteo2005}. The Eddington ratios found with this modification (model IIIb) are typically lower than in model IIIa. Again we set $f_{\rm Edd}>10^{-3}$.  Model IIIb is therefore representative of a simple attempt to decrease further the typical accretion rate of MBHs at low redshift.

The main assumptions regarding the dynamical evolution of the MBH population in our models can be found in \cite{VHM, Volonterietal2005,Vrees2006}.  

\section{Luminosity functions}

We have calculated the luminosity functions by implementing the different accretion models within a comprehensive model for black holes evolution in a Cold Dark Matter universe. The history of dark matter halos and their associated black holes is traced by merger trees (Volonteri, Haardt \& Madau 2003). The evolution of the massive black hole population traces the accretion and dynamical processes involving black holes. We have assumed that accretion onto nuclear black holes is triggered by halo mergers, and we have then computed the accretion rate and luminosity of the active systems as described in the previous section. At every step of the simulations we apply the appropriate Eddington rate to accreting MBHs. For model I  $f_{\rm Edd}=1$ for all MBHs at all times. For model II, $f_{\rm Edd}$ is only redshift dependent (see Eq. \ref{eq:shankar2 }), while for models  IIIa and IIIb we determine $f_{\rm Edd}$ as a function of the black hole mass at the beginning of the timestep, and of the host velocity dispersion. 

The luminosity functions are computed selecting the black holes which are active at the chosen output redshifts ($z=0.5, 1, 2, 3$), and weighting each of them according to the Press \& Schechter function.  We derive the AGN bolometric luminosity as 
$L=\epsilon f_{\rm Edd}(t)  \dot{M}_{\rm Edd} c^2$. We apply the  bolometric corrections and spectrum of \cite{Marconietal2004} to model the SED in the blue band. 
The spectrum of unabsorbed (here dubbed Type I) AGN is described by a  power--law with photon index $\Gamma=-1.9$, exponentially cut--off at $E_c=500$ keV. The averaged SED of absorbed (Type II) sources (i.e., sources with absorbing column $\log{(N_{\rm H}/{\rm cm}^{-2})}>22$) is described by the same Type I spectrum for $E> 30$ keV, and by a power--law (continuously matched) with photon index  $\Gamma=-0.2$ \citep{Sazonov2004} at lower energies. 
The  Type II/Type I ratio is, in general, function of luminosity and redshift. 
Here, we adopt the model \#4 of \cite{lafranca2005}, which explicity allows for 
the redshift and luminosity evolution of the $N_H$ distribution, providing the best 
fit model to the HXLF of the HELLAS2XMM sample \citep{fiore}.
Error bars, at 1--$\sigma$, for the theoretical LFs have been computed assuming Poisson statistics. 

\begin{figure}   
   \includegraphics[width=8cm]{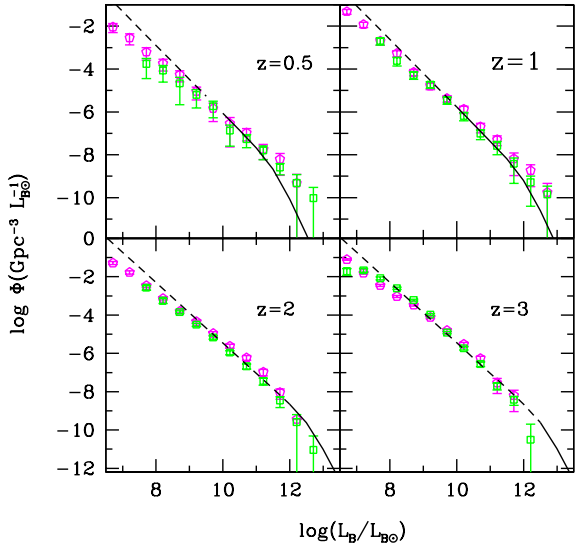} 
%      	\vspace{0.5truecm}
   \caption{Luminosity function of AGN in the B-band corrected for absorption. Clockwise: z=0.5, 1, 2, 3. Green squares  show model I ($f_{\rm Edd}$=1),  magenta pentagons model IIIa  ($f_{\rm Edd}$ luminosity dependent).  {\it Solid lines :} 2QZ/6QZ LF. The dashed lines show the extrapolation to faint magnitudes. }
   \label{fig1}
\end{figure}

\begin{figure}   
   \includegraphics[width=8cm]{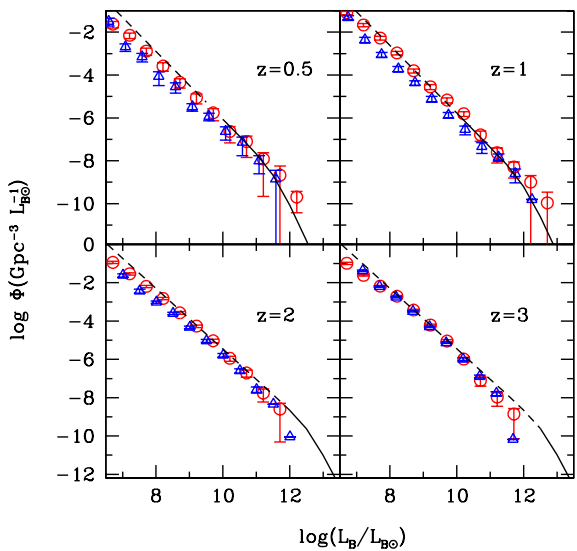} 
%   	\vspace{0.5truecm}
   \caption{Luminosity function of AGN in the B-band corrected for absorption. Clockwise: z=0.5, 1, 2, 3. Blue triangles  show model II ($f_{\rm Edd}$ redshift dependent),  red circles model IIIb  ($f_{\rm Edd}$ luminosity dependent).}
   \label{fig2}
\end{figure}

We have compared our theoretical OLF at different redshifts to the OLF by \cite{croom2004} obtained by merging the 2dF QSO Redshift Survey (2QZ), with the 6dF QSO Redshift Survey (6QZ).  Figures \ref{fig1} and \ref{fig2} show an absorption-corrected OLF against the best fit models by \cite{croom2004}.  In order to guide the eye, we have extrapolated the OLF at the faint end below the observational limit (dashed lines) adopting the same \cite{croom2004} fit. We note, however that the theoretical OLF flattens towards lower luminosities, as has been observed \citep{richards2005, hunt2004}.

The agreement between the theoretical and observed OLFs is good for all models (I, II, III), which result almost undistinguishable  in the observed luminosity range (solid line in Figures \ref{fig1} and \ref{fig2}).  
Model I overpredicts the bright-end of the OLF at low redshift ($z<1$) more substantially than models II and III; on the other hand models with subEddington accretion tend to underpredict the bright end at high redshift. In fact, we cannot form MBHs massive enough to power bright AGN by $z=3$ if models II or III are considered at all times\footnote{We have modified model III, assuming $f_{\rm Edd}$=1 at $z>12$ in order to obtain significant growth of MBHs before z=6. See \cite{Vrees2006} for a discussion on the constraints set by the OLF of bright AGN at $z=6$.}.

The  OLF is known to be a biased tracer of the accretion history of
MBHs, missing the vast majority of Type~II objectes. Moreover it spans a 
smaller range in luminosity compared to the HXLF \citep{ueda2003}. 

\begin{figure}   
   \includegraphics[width=8cm]{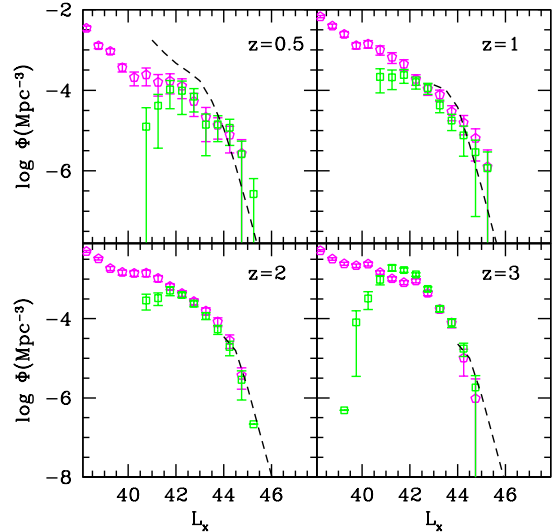} 
   	\vspace{-0.5truecm}
  \caption{Luminosity function of AGN in the hard X-ray band [2-10 Kev]. 
Symbols as in Figure 1. The dashed lines show the 
Ueda et al. (2003) HXLF in the observationally constrained luminosity range.}
   \label{fig3}
\end{figure}

\begin{figure}   
   \includegraphics[width=8cm]{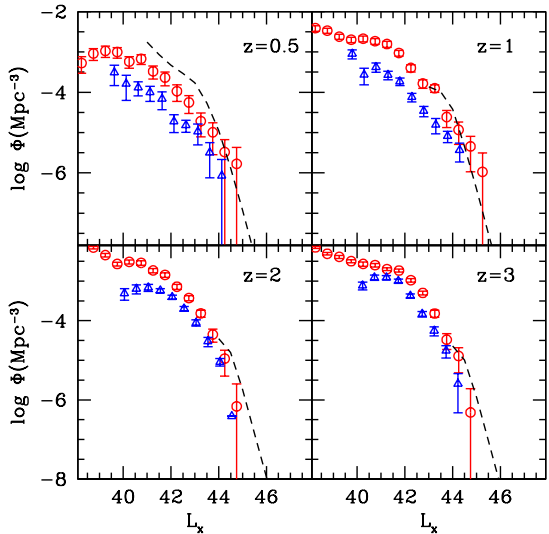} 
   	\vspace{-0.5truecm}
   \caption{Luminosity function of AGN in the hard X-ray band [2-10 Kev]. Symbols as in Figure 2. The dashed lines show the Ueda et al. (2003) HXLF in the observationally constrained luminosity range.}
   \label{fig4}
\end{figure}

When we compare the theoretical and observed HXLFs of unabsorbed AGN (Figures \ref{fig3} and \ref{fig4}),
large differences at the faint end become apparent. Model I largely underestimates the faint end
 at $z\la1$, while at $z\ga2$ it agrees very well in the luminosity 
range probed by current surveys. Model II underestimates the normalization of  HXLF at all redshift, although the shape is satisfactorily matched. In model II accretion is simply not enough to grow black holes massive enough to account for the bright end of the HXLF. If the normalization is changed from, $f_{\rm Edd,0}=0.3$ to $f_{\rm Edd,0}=1$ (cfr. Lapi et al. 2006), the model fares much better at high-z (basically corresponding to model I at $z>2$), but incurs in the same issues of model I at lower-z.

Models IIIa and IIIb fare better in reproducing the low redshift HXLF, but predicts a large 
population of faint AGN at $z\ga2$ and slightly underestimate the bright end at high redshift ($z>2$). 
We note here that our approach differs from that by \cite{hopkins2005b}. The starting point of \cite{hopkins2005b} is the HXLF, from which they derive the quasar birth rate, and consequently the OLF and other diagnostics. Our approach instead follows the evolutionary path of MBHs and AGN, that is the population evolves self-consistently along the cosmic epochs, according to the accretion properties stated in \S 2.  The HXLF becomes therefore a constraint, rather than an input of the model as in \cite{hopkins2005b}.

We have investigated the impact of the initial conditions, by applying model IIIa to a scenario
in which seeds are much more massive, as in \cite{kous}. The resulting luminosity functions are negligibly different with respect to the corresponding models assuming smaller seeds. This is because observable properties are mainly determined by the accretion history rather than by the initial conditions.  Differences arise only at luminosity around $10^{42}$ erg s$^{-1}$, as a bump at $10^{42}$ erg s$^{-1}$, where the MBHs with a mass around that of the initial seeds are clustered, and a sharp decrease faint-ward of $10^{42}$ erg s$^{-1}$.

\section{Faint X-ray counts}
The luminosity functions (OLF and HXLF) are the most sophisticated analysis of the evolution of AGN as a function of luminosity and redshift. On the other hand the available surveys do not probe yet the extreme faint end where theoretical models mostly differ, except at very low redshift. Figures \ref{fig3} and \ref{fig4} show that theoretical models predictions, at $z> 0.5$ branch off at luminosities not yet sampled by the HXLF. Number counts are the results of integrating over intrinsic luminosity and distance. Number counts are a weaker test than the luminosity function, as AGN with a wide range of intrinsic luminosities are included at each flux. Nevertheless, they are the most direct probe of the AGN population. They are independent of cosmology, allow to probe further the faint population, where the HXLF is still prohibitive because of spectroscopic flux limits. We therefore compute the expected X-ray counts for the same AGN population that we used to determine the luminosity functions and compare the model results to the most recent determinations of X-ray counts and their redshift distribution. 
\subsection{Basic Equations}

The number of sources (per unit solid angle) seen in the flux range $S$, $S+dS$ 
by an observer located at $z_0$, is

%%%%%%%%%%%%%%%
\begin{equation}
\frac{dN}{d\Omega dS}(z_0,S)=\int^{\infty}_{z_0}
\left( \frac{dV_c}{dz d\Omega} \right) n_c(z, S)  \, dz,
\end{equation}
%%%%%%%%%%%%%%%

\noindent
where $dV_c/dz d\Omega$ is the comoving volume element per unit redshift per
unit solid angle, and $n_c(z,S)$ is the comoving density of sources at
redshift $z$, with observed flux in the range $[S, S + dS]$.
The integrated flux of a source observed at $z_0$ is given by
%%%%%%%%%%%%%%%%%
\begin{equation}\label{eq:flux}
S=\frac{1}{4\pi \, d_L^2(z_0,z)}\int_{\Delta \nu} \tilde L_\nu (M) d\nu ,
\end{equation}
%%%%%%%%%%%%%%%%%
\noindent
where $\nu=\nu_0 (1+z)/(1+z_0)$, 
$d_L(z_0,z)$ is the luminosity distance between redshift $z_0$ and $z$, 
$\tilde L_\nu(M)$ is the 
specific luminosity {\it averaged over the source lifetime}  
(assumed to be only a function of the BH mass, $M$), and $\Delta \nu$ is
the {\it rest-frame} frequency bandwidth. 

The background specific intensity $J_{\nu_0}(z_0)$ observed at redshift 
$z_0$ at frequency $\nu_0$, is 
%%%%%%%%%%%% 
\begin{equation}\label{eq:J}
J_{\nu_0}(z_0)= \frac{(1+z_0)^3}{4\pi}\int^{\infty}_{z_0}
\epsilon_\nu(z)  \frac{dl}{dz}dz,
\end{equation}
%%%%%%%%%%%% 
where $dl/dz$ is the proper line element, and the comoving specific emissivity 
$\epsilon_\nu(z)$ is given by
%%%%%%%%%%%%
\begin{eqnarray}\label{eq:eps}
\epsilon_\nu(t)&=&\int dM\,\, \int_0^t L_{\nu}(t-t',M)\frac{dn_c}{dt'dM}dt' \nonumber \\
&\simeq& \int dM\,\, \tau \tilde L_{\nu}(M) \frac{dn_c}{dt dM}. 
\end{eqnarray}
%%%%%%%%%%%%
The second approximated equality holds once we consider the  
source light curve averaged over the typical source lifetime $\tau$,  
assuming the formation rate of sources per unit mass as constant over such timescale.

\subsection{Number counts}

\begin{figure}   
   \includegraphics[width=8cm]{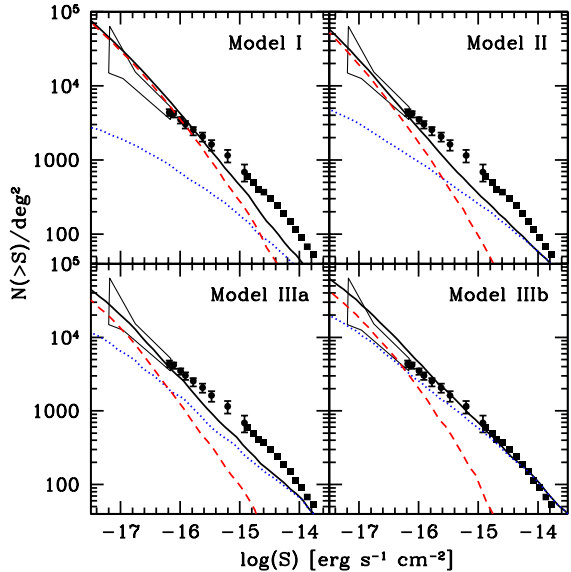} 
%   	\vspace{-1.5truecm}
%   	\vspace{1.5truecm}
   \caption{Predicted $\log(N)/\log(S)$ in the observed soft--X band 
[0.5--2 keV] for the different models. Dotted lines show the contribution
of sources with $z<2$, whereas dashed lines the contribution of sources
with $z>2$. Solid line is the sum of the two components. Points
report data obtained with {\it Chandra} (dots; Moretti et al. 2003) and 
{\it XMM} (squares; Baldi et al. 2002), and the bow-tie indicates the result of the
fluctuation analysis of the {\it Chandra} deep field \citep{Bauer2004}.
}
   \label{fig5}
\end{figure}

\begin{figure}   
   \includegraphics[width=8cm]{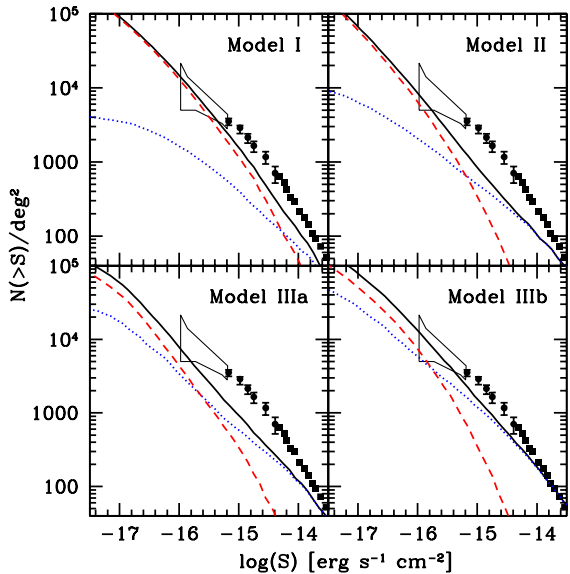} 
%   	\vspace{-1.5truecm}
 %  	\vspace{1.5truecm}
   \caption{Predicted $\log(N)/\log(S)$ in the observed hard--X band 
[2--10 keV] for the different models. Lines and points are the same of
Fig.~\ref{fig4}.
}
   \label{fig6}
\end{figure}

We computed the number counts predicted by our different assumptions concerning the 
evolution of the Eddington--parameter along the cosmic history. 
The soft and hard X--ray logN/logS are shown in fig.~\ref{fig5} and fig.~\ref{fig6}, respectively. 
The total counts (solid lines) are divided into the contribution of sources at $z<2$ (dotted lines), and $z>2$ (dashed lines). 
Model results are compared to a compilation of
X--ray data from the {\it Chandra} (dots, \cite{Moretti2003}), and {\it XMM} (squares; \cite{Baldi2002}) deep field surveys. 
The bow--tie indicates results of the fluctuation analysis of the {\it Chandra} 
deep field \citep{Bauer2004}.

Model I fails to reproduce the slope of the observed logN/logS,  
falling short in the number of bright objects, and slightly overpredicting faint AGN. 
Moreover, the counts are dominated by high redshift sources 
for fluxes below $\log S<-14.6$. 
In Model~II, the redshift
distribution of AGN is somewhat pushed towards lower redshift, because of the relatively longer
accretion time involved. The model underpredicts the counts both
in the soft and hard X--ray bands, as BHs do not have enough time to grow.
Model IIIa matches well the observed logN/logS in the soft band, 
but underpredicts the counts in the hard band. 
Finally, Model~IIIb gives a reasonable good description in both bands, 
though it slightly overpredicts counts at very faint fluxes. 
AGN number counts are dominated, in the entire observed
flux range, by low redshift objects, the contribution of sources at $z>2$ 
becoming significant only at fluxes as faint as the limits of the more recent surveys.

\subsection{Redshift distribution of X--ray selected AGN}

\begin{figure}   
   \includegraphics[width=8cm]{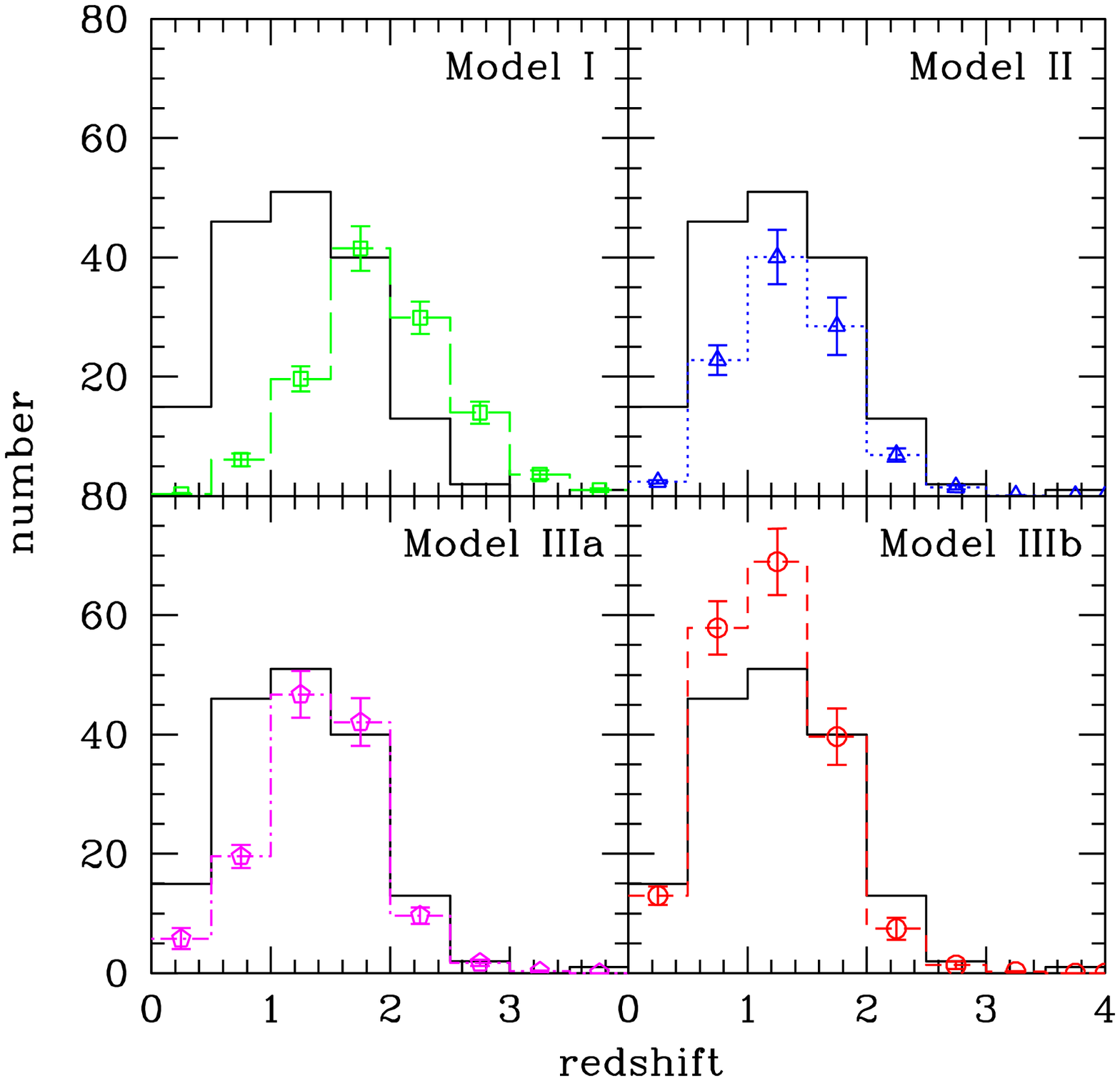} 
%   	\vspace{-1.5truecm}
   \caption{Distribution of Type~I AGN as function of redshift for different
models (points) compared to the result of Eckart et al. (2005; solid line)
from the SEXSI program. In order to compare our results to
 the observed distribution, we have convolved the predicted number counts
with the sky coverage of the survey at different flux limits 
\citep{Harrison2003}, and we have split the data so to match our definition
of Type~I/Type~II sources.}
   \label{fig7}
\end{figure}

Aiming at constraining further the 4 different models employed, we compare the predicted
redshift distributions to the results of the {\it Serendipitous Extragalactic 
X-ray Source Identification} (SEXSI) program, a survey designed to resolve 
a large fraction of the 2--10 keV cosmic X--ray background \citep{Eckart2006}. The survey covers 1 deg$^2$ for fluxes $>1\times 10^{-14}\;\Xflux$, and 2 deg$^2$ for fluxes 
$>3\times 10^{-14}\;\Xflux$. Given the large survey area, the SEXSI program minimizes the effects of cosmic variance. The catalog contains a total of 
477 spectra, among which 438 have redshift and optical  identification
\citep{Eckart2006}. 
The Type~I AGN redshift distribution of the SEXSI program and of our
selected models are shown in fig.~\ref{fig7}. 
Note that we splitted the original SEXSI
data in order to match our definition of Type~I/Type~II sources. 
Moreover, we have convolved the predicted number counts
with the sky coverage of the survey for different flux limits 
\citep{Harrison2003}. 

Model I fails completely to reproduce
the observed redshift distribution. In particular, the model largely underestimates 
the number of sources at $z<1.5$: the distribution peaks at redshift higher than observed. 
A better match to data is achieved by Model II.  
The general shape of the distribution is reproduced, although the model largely 
underestimates the total number of sources, as already pointed out. 
Model IIIa and IIIb are in reasonable agreement with the data. 
Model IIIa falls short to the data at $z<1$, while 
Model IIIb overpredicts the number of sources observed in the range $0.5<z<1.5$. 
In conclusion, the best agreement with the
observed logN/logS, and with the redshift distribution of sources is
found assuming a luminosity dependent Eddington rate \citep{hopkins2005a}. 

We also tested that, in a model in which BH seeds are more massive, as in \cite{kous}, the
resulting $\log N/\log S$ and redshift distribution do not differ 
significantly with respect to models with earlier, smaller seeds. 
In conclusions, different formation
scenarios for seed BHs are difficult to discriminate on the basis of 
available X--ray deep field surveys.

%
%\begin{figure}   
%   \includegraphics[width=8cm]{red_dist_gt22} 
%   	\vspace{-1.5truecm}
%   \caption{Same of Fig.~\ref{fig6} but for Type~II AGNs. {\bf Farei la figura SENZA i dati, oppure leviamola del tutto.}}
%   \label{fig7}
%\end{figure}

\subsection{Unresolved X--ray Background}

According to \cite{Moretti2003}, the intensity of the total X--ray 
background (XRB) is  $7.53\pm 0.35 \times 10^{-12}$ and 
$2.02\pm 0.11 \times 10^{-11}\;\Xback$ in
the 0.5--2 keV, and 2--10 keV energy bands, respectively. 
A large fraction, $\simeq 94$\%, of the soft XRB (SXRB) has been 
attributed to sources with fluxes exceeding $2.4\times 10^{-17}\;\Xflux$, 
while $\sim 89$\% of the hard XRB (HXRB) is resolved into sources 
whose flux is $\geq 2.1\times 10^{-16}\;\Xflux$ (Moretti et al. 2003). 
More recently, \cite{Hickox2005} estimated the unaccounted fraction
of the XRB due to extragalactic unresolved sources as 
$1.77\pm 0.31\times 10^{-12}\;\Xback$ in the soft X--ray energy band 
(0.5--2 keV) and $3.4\pm 1.7\times 10^{-12}\;\Xback$ in the hard X--ray 
energy band (2--8 keV).

\begin{figure}   
   \includegraphics[width=8cm]{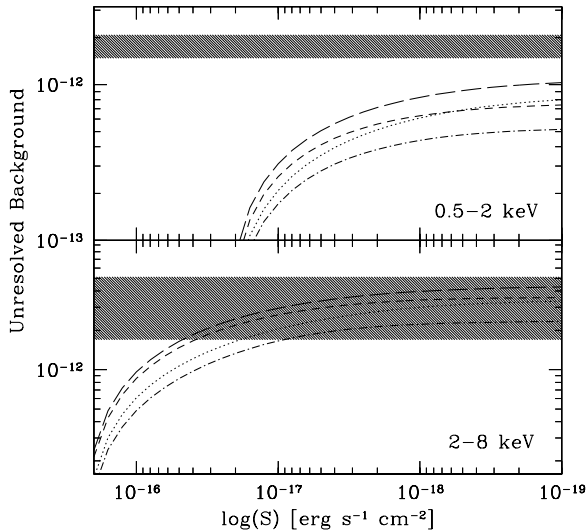} 
%   	\vspace{-1.5truecm}
   \caption{Predicted cumulative contribution to the unresolved XRB from 
different models as function of the X--ray flux.
Different lines refer to different model: Model I (long-dashed line), Model II
(dotted line), Model IIIa (dot-dashed line), and Model IIIb (short-dashed line). The shaded area show the measured unaccounted background as reported by
Hickox \& Markevitch (2005). Top panel: unaccounted XRB in the observed soft--X
band [0.5-2 keV]. Bottom panel: unaccounted XRB in the observed hard--X band
[2--8 keV].}
   \label{fig8}
\end{figure}

Using our different models, we compute the contribution to the unresolved XRB 
due to faint AGN lying below 
sensitivity limits of current X--ray surveys. The cumulative contribution
from sources with flux above a given threshold is shown in Fig.~\ref{fig8}, where it is  
compared to the recent estimate of Hickox \& Markevitch (2005; shaded area).
Different line styles refer to different models: 
Model I (long-dashed line), Model II (dotted line), Model IIIa (dot-dashed 
line), and Model IIIb (short-dashed line). The unresolved SXRB and HXRB are shown in 
the top and bottom panel, respectively.
Although significant differences (within a factor of $\simeq 2$) are found between different models, 
all predict a contribution to the unresolved XRB 
consistent with available limits. 
Note that all models can account for the whole unresolved HXRB, while they give at most 50\% of the unresolved SXRB. 
Our results imply the existence of a further population of faint X--ray sources in the soft band. 
Indeed, Salvaterra et al. (2006) 
have found that a significant contribution to the unresolved XRB may come 
from accreting BHs at very high redshift ($z>6$), a redshift range not considered here. 
By means of a dedicated
model of the SMBH assembly at early times, consistent with the SDSS
OLF at $z=6$ and with ultra--deep X--ray constraints, they found that 
the contribution to the SXRB of very high redshift, undetected AGN could 
be as high as $\sim 0.4\times 10^{-12}\;\Xback$, providing the residual unresolved flux. 
The contribution of such population to the HXRB is still consistent with the
available limits, being only $\sim 0.9 \times 10^{-12}\;\Xback$. 

\section{Discussion}
We have attempted in this paper to place constraints on the global accretion properties of the MBH population at $z<3$.  We consider the full cosmological evolution of MBH embedded in their host halos, rather than adopting an empirical approach which takes as a starting point the observed LF in a given band in order to explain the properties of AGN in other bands. We focus here on the strength of accretion, parameterizing the accretion rate as a function of the Eddington rate, $f_{\rm Edd}$. We show that simple models which assume $f_{\rm Edd}=1$ (model I), although highly idealized, still are able to explain satisfactorily the growth of MBH, as traced by the OLF, in a large redshift range, as claimed previously by various investigations.  Even at low redshift, where hierarchical models start to fail in their predictions, the OLF is very well reproduced.  The HXLF probes fainter sources than the OLF, and the simplistic model I is shown to provide a poor match with the HXLF at $z<1$.  A decreasing average Eddington ratio \citep[model II]{Shankar2004} provides a better agreement with the \emph{shape} of the HXLF, but underestimates the \emph{normalization}, as there is not enough time to grow high mass MBHs if the Eddington ratio is not large at $z>3$.  A model (model III) with an Eddington ratio depending on luminosity \citep{hopkins2005a, Shankar2004} seems to be the best match at low redshift. The difference among the models is magnified when the faint X-ray counts, and in particular their redshift distribution, are calculated. Model I has a strictly hierarchical growth of MBHs, and moreover, a univocal relationship between MBH mass and AGN luminosity. Model II allows massive black holes at low redshift to shine at lower luminosity, as the accretion rate, in units of the Eddington one, decreases with time. A rigid redshift dependence, however, creates two problems: first, subEddington accretion at $z>3$ implies a much slower growth for high redshift AGN, thus causing an underestimate of the LFs \emph{normalization} at all the considered redshifts. Second, the evolution of the Eddington rate is not fast enough, at $z<3$, to account for the faint end of the LFs, in particular the HLF.  The accretion rates, and Eddington ratios, predicted by simulations are here embedded into a cosmological framework (models IIIa and IIIb). The resulting AGN population provides a satisfactory match with  the OLF, HXLF and faint X-ray counts at low redshift, while sources at high redshift are more problematic. The low accretion rates predicted by simulations imply very long growth timescales for black holes, and therefore underestimate the occurrence of  bright quasars powered by billion solar masses black holes at high redshift ($z>2$). 

It is not clear, however, if the accretion rate found in simulations, which relates to model III, i.e., strongly subEddington for low-luminosity sources (and therefore in all cases for small MBHs) indeed applies at very high-redshift.  The simulations on which \cite{hopkins2005a} model is based assume mergers between "normal" galaxies. How "normal" are galaxies at $z>3$?  It might indeed be possible that the conditions in pre-galactic structures at high-redshift (e.g., the disturbed morphological state of galaxies) cannot be studied with simulations of mergers of evolved discs and bulges, but would require different initial conditions for the merging galaxies.  Future deep surveys can help us distinguish between various models in the high redshift Universe (see Figure \ref{fig3}), and can probably locate the time, if any, for a transition between messy mergers with on average efficient accretion onto the MBHs, and standard galactic mergers, predicting long periods of inefficient accretion.

Finally, we computed the contribution to the XRB of faint AGN lying below the
sensitivity limits of current X--ray surveys. We found that all model predict
a contribution to the unresolved XRB consistent with the available limits,
accounting for the whole unresolved XRB in the 2--8 keV band and for 
$\sim 50$\% in the 0.5--2 keV. The residual background intensity in the soft
band may be provided by AGN shining at $z\ga 6$ (Salvaterra et al. 2006).

\section{Acknowledgments}
We would like to thank Dave Alexander for interesting discussions and helpful comments. This research was supported in part by the National Science Foundation under Grant No. PHY99-07949. NSF-KITP-06-88 pre-print.

\bibliographystyle{mn2e}
\bibliography{xrc.bib}

\end{document}